# Two Slits Interference Is Compatible with Particles' Trajectories

Vladimir V. Kisil

October 26, 2018

## Abstract

We propose a simple numerical experiment of two slits interference of particles. It disproves the popular belief that such an interference is incompatible with a knowledge which slit each particle came through or, more generally, "quantum particles could not have trajectories". Our model is an illustration to the contextual interpretation of quantum probabilities.

> It is all mysterious. And the more you look at it the more mysterious it seems.
> Richard Feynman, [4, § 1–5].

## 1 Introduction

There is a recent interest in revisions of foundations of quantum mechanics and its relation to classic one. An example of such fundamentals is the double slits interference of electrons, which usually opens quantum mechanical textbooks [4]. A consideration of that thought experiment (an actual interference could not be realised on just two slits—one need a crystal lattice instead) is a common conclusion that "electrons could not move along definite paths". Besides that vague statement the experiment is used for s derivation of the uncertainty principle and a justification of quantum rules for adding probabilities.

Recently those rules were analysed from the viewpoint of the *contextual probability*, see [2, 7, 8], papers in this volume and references herein. It

---

allows to explain a difference between classic and quantum rules for addition of probabilities given by identities:

$$P_{12} = P_1 + P_2 \qquad \text{and} \qquad (1)$$
$$P_{12} = P_1 + P_2 + 2\sqrt{P_1 P_2}\cos\theta \qquad (2)$$

correspondingly. The contextual suggestion is to consider a probability $P$ of an outcome $A$ depending from the context $S$ of experiment, the contextual probability is denoted by $P(A|S)$. For example, in the two slits experiment one may assume up to three different contexts: $S_{12} =$ both slits are open, $S_1 =$ only the first slit is open, and $S_2 =$ only the second slit is open. Then in the above formulas (1)–(2) probabilities should be understood as follows:

$$P_{12} = P(A, S_{12}), \qquad P_1 = P(A, S_1), \qquad P_2 = P(A, S_2).$$

In the classical situation the context $S_{12}$ is the disjoint union of two contexts $S_1 \cup S_2$ and classic addition formula (1) could be rewritten as $P(A|S_{12}) = P(A|S_1) + P(A|S_2)$. In the quantum case the third term in the formula (2) reflects the change of context (see [8] and Section 4 for details) between different experimental settings, e.g. if different slits are open. This allows to wipe out any mystery from quantum formula (2) on the theoretical level.

On the other hand the generality of that construction obscures its connection with concrete experiments. It is desirable from psychological and pedagogical point of view to have a toy model based on contextual principles which could imitate double slit interference, for example. We describe such a model in this paper and arrive to the conclusion that *interference of particles is perfectly compatible with the precise knowledge of their trajectories.* It could appear to be disputing with the dominant Copenhagen Interpretation of quantum mechanics. But we will argue in Remark 3 that in facts our model is even more in line with the Copenhagen philosophy than those typical explanations referring to mysterious particles-without-a-path.

## 2 Scheme of the Experiment

Let us describe the scheme of our experiment. There is no claim that common words like "atom", "electrons", and "spin" in the text bellow exactly correspond to a physical reality, merely they are chosen for illustration purposes only. The scheme is graphically represented on the Figure 1. There is a source of particles, which emits them in the horizontal direction with a uniform distribution over a vertical segment. On their way particles meet two slits, which are named "1" and "2" and could be open or closed in an arbitrary combination. A particle coming to a closed slit disappears without a trace.

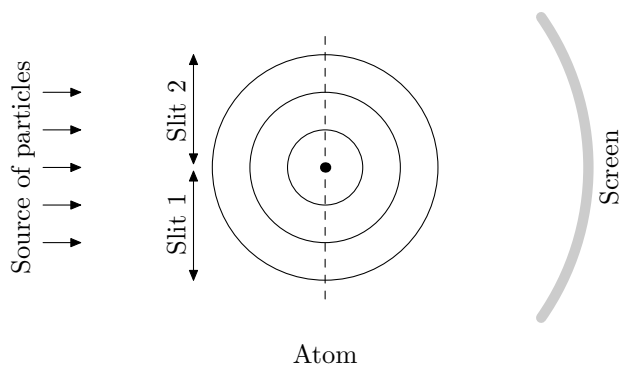

Figure 1: Graphical representation of the model.

Particles coming through an open slit will interact with an atom. The atom have equidistant orbits. The *principal ingredient* of the model is that each orbit have a spin, which could take exactly two values named again as "1" and "2". An interference of the particle is presented by a displacement in the vertical direction without an alteration of its velocity. It could occur *if and only if* the name of spin for the nearest orbit is the same as the name of slit the particle came through. If an interference did happen than the spin of that orbit *will be changed* to the opposite. This is exactly the place their our model shows a *contextual behaviour*. Indeed if only one slit is open then each orbit could produce at most one interferential displacement. However if particles come through both open slits then a rich interference appears due to randomness in their emission.

Besides an interference all particles also experience a scattering on the atom as a whole. This manifests itself in changes of their directions of motion and is observed on a semicircular screen behind the atom. We chose the rules for interference and scattering to be simple yet reproducing the popular pictures from the quantum mechanical textbooks [4, Fig. 1–3], [11, Fig. 5.2].

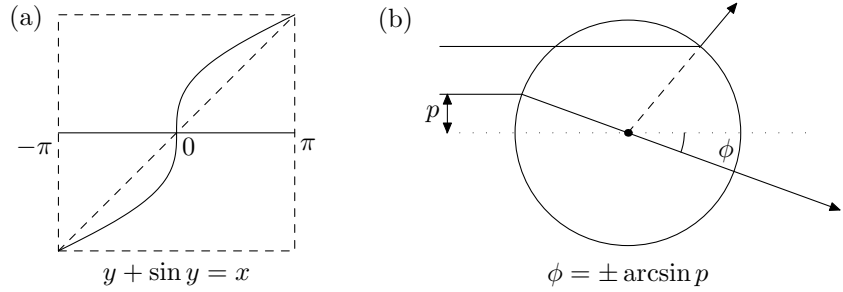

Figure 2: The rules for interference (a) and scattering (b).

**Convention 1** Our assumptions are as follows (see Figure 2):

(a) An *interference displacement* (if occurs at all) is directed out of the nearest orbit to the particle when it crossed the dashed line on the Figure 1. The amount of displacement comes from the formula:

$$y + \sin y = x, \qquad (3)$$

where $x$ and $y$ are distances to the nearest orbit before and after interference correspondingly both measured in units equal $\pi*$(the distance between orbits). See Figure 4(a) for the graph of that function.

(b) A particle *scatters* with probability $1/2$ either from the front or the back semicircular boundary of the atom. The direction of scattering forms a sharp angle with the original horizontal direction and belong to the radius of circle coming through the point of their contact, see Figure 2(b). Analytical expression of that configuration is:

$$\phi = \pm \arcsin p. \qquad (4)$$

After scattering the particle is registered by a semicircular screen which is sufficiently far away from the atom to neglect the atom size.

REMARK 2 Note that the only non-deterministic elements enter the scheme are: (i) the initial choice of the position of the particle, and (ii) the choice of $+$ or $-$ in the formula (4). Those two choices are independent each other and we could associate them both with the process of the particle emission. Apart from them the rest of algorithm of interference-scattering and registration of particle is completely deterministic. We also able to keep information about the slit the particle come through at the moment of its registration on the screen.

It is easy to realise the above scheme on many computer languages. We chose a programming of the MuPAD [3] open computer algebra system because it provides a good graphical presentation of the obtained results and has a free license for a usage within academia. Therefore to reproduce and verify our results anyone does not have to own an expensive licence for the commercial software. The complete listing of the used program together with some comments is given in Appendix A.

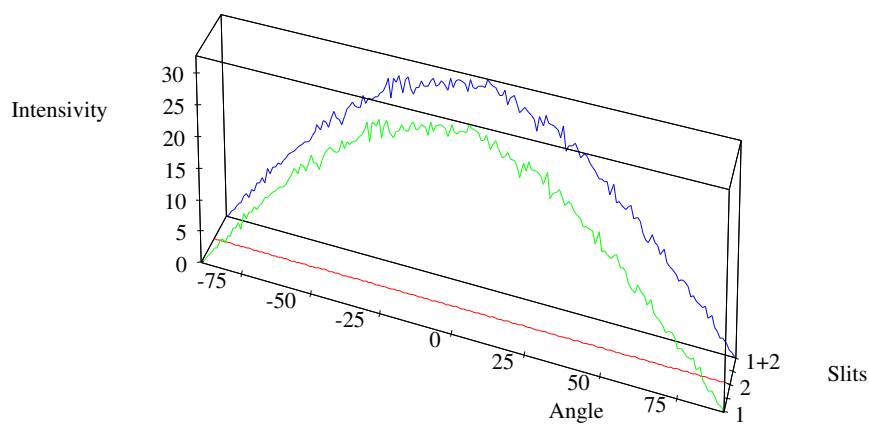

Figure 3: An absence of interference if only one slit is open.

## 3 Obtained Results

The described model could be tested in two essentially different contexts: with just one skit or both slits open. The results of these two tests are shown on Figures 3 and 4 and will be explained now. The output consists of the three graphs, which show the number of particles for each degree on the screen coming through first, second, and both slits correspondingly. Because these graphs could be identical or very close we put them in perspective on three parallel planes and coded them by colours. The nearest (green) graph shows particles coming through the first slit, the second (red) is drawn for the second slit, and the third (blue) graph is the sum of the both previous graphs.

If just the first slit is open (see Figure 3) then the first graph is essentially affected only by scattering and did not shows any visible signs of interference; the second graph is identically zero; and therefore the third graph is equal to the first one. This is exactly that we know from textbooks [4, part (b) of Figs. 1–1, 1–2, 1–3] for a behaviour common to both quantum and classic mechanics. The included program produces that output if experimenter sets in its beginning the following values of variables:

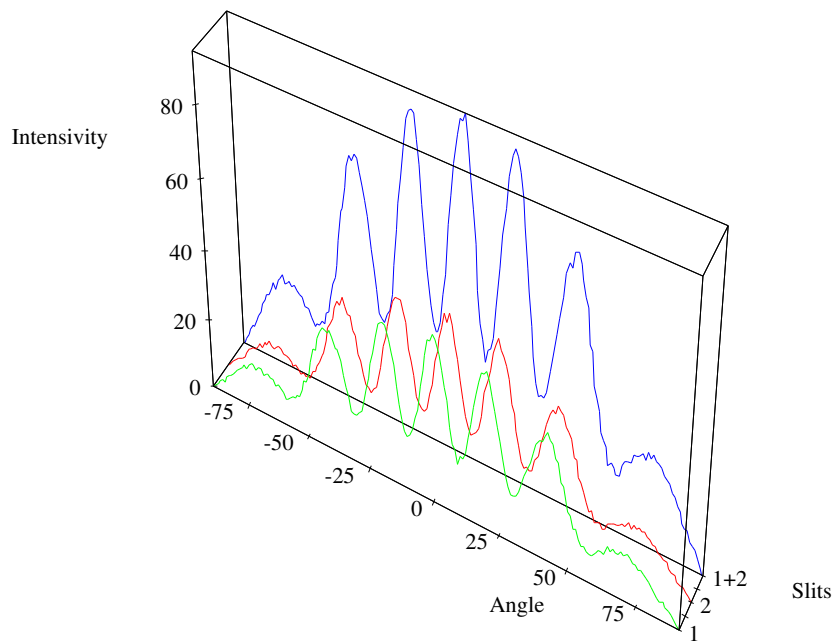

Figure 4: The presence of interference if both slits are open.

first_slit_is_open := **TRUE**:
second_slit_is_open := **FALSE**:

On the other hand, if both slits are open and particles will come through them in a random order then the context of the experiment will be different and interference will affect particles coming through both slits. Such an outcome is shown on Figure 4: the both 1st (green) and 2nd (red) graphs are almost identical and the total sum is like a double of any of them. This is in a good agreement with the quantum interference of particle illustrated for example in [4, Fig. 1–3], [11, Fig. 5.2]. The included program produces that output if values of both variables are set to be "TRUE" (the actual values in the listing given in Appendix A):

first_slit_is_open := **TRUE**:
second_slit_is_open := **TRUE**:

Note that if an experimenter changes the context again and will firstly send a half of particles (still randomly distributed) only to the first slit and afterwards another half to the second slit then interference will not occur again—just like in the case of one open slit. This is similar to the thought experiment with "watching electron", which destroys the interference completely [4, § 1–6]. In that way our model reproduces one more feature associated with mysterious quantum behaviour.

## 4 Do Particles Have Trajectories?

Let us recall the "Proposition A" from of the popular textbook [4, § 1–5]:

**Proposition A** Each electron *either* goes through slit 1 *or* it goes through slit 2.

Unfortunately the another important assumption was not highlighted by the authors of [4] so explicitly:

**Proposition B** The distribution of electrons coming through slit 1 is the same *regardless* either slit 2 is open or not.

This two assumptions together imply the addition rule $P_{12} = P_1 + P_2$ (1), which is wrong in the quantum case: we could not get the third joint (blue) distribution on the Figure 4 adding together two identical (green) distributions from the first slit on Figure 3. Because Proposition B was hidden in [4] the only responsible for that failure seems to be Proposition A:

> ...undoubtedly we should conclude that *Proposition A is false*.
> It is *not* true that the electrons go *either* through hole 1 or 2.
> ([4, § 1–5], emphasis of the original.)

But as we saw in the previous section Proposition A is perfectly compatible with the interference of electron. Instead we would blame Proposition B for the failure of an accurate prediction. This proposition is explicitly *non-*contextual, it supposes that an outcome of a particular event is independent from the whole context. In contextual framework we drop the Proposition B and get in the terms of contextual probabilities:

$$P(E_1|S_1) \neq P(E_1|S_{12}), \qquad \text{and} \qquad P(E_2|S_2) \neq P(E_2|S_{12}),$$

where $E_i$ are the events that an electron come through slit $i$, $i = 1, 2$, and $S_1$, $S_2$, and $S_{12}$ are contexts where only slit 1, only slit 2 is, and both slits are open. Then instead the definitely wrong statement $P(E_{12}|S_{12}) = P(E_1|S_1) + P(E_2|S_2)$ the Figure 4 represent a true contextual addition of probabilities:

$$P(E_{12}|S_{12}) = P(E_1|S_{12}) + P(E_2|S_{12}). \tag{5}$$

REMARK 3 It is interesting to note that our model for interference should satisfy a most orthodox follower of the Copenhagen Interpretation. Indeed in that interpretation the measuring apparatus is a part of the measured system and the *uncertainty principle* is a consequence of impossibility to make its backreaction to the system arbitrary small. From that point of view the two slits are also parts of the system and their influence on an outcome of experiment could not be neglected. In our model slits preserve some information about electrons coming through them. This provides a device for an indirect interaction between electrons even if any direct interaction is excluded.

It could be interesting to construct more advanced and realistic models based on the same principle as our one. For example, we can study an interference on a regular lattice of simple atoms without a complicated internal structure. In that way it is reasonably to expect that a collective behaviour of atoms in the lattice will allow to get consequences similar to our Convention 1 starting from simpler and less artificial assumptions.

Another interesting direction for a research is as follows. We confirmed a correct form (5) for addition of quantum probabilities which is similar to the classic one (1), both of them are just different realisations the same general contextual formula. Similarly it was proposed recently [10] to obtain both quantum and classic brackets from the same common source—the *p*-mechanical brackets. It may be possible and promising to combine both approaches (contextual probability and *p*-mechanical brackets) in order to wipe out an unnatural opposition of quantum and classic worlds.

# A   Appendix: the Listing of the MuPAD Program

Here is the listing of the program for the MuPAD [3] software to create Figures 3 and 4. The listing is typeset with the help of the *free software* Lgrind [1].

The code is short but you do not need even to retype it to use for our own numerical experiments. If you get the source of this article [9] form the arhiv.org and type the command

```
latex 0111094.tex
```

then the file interfr1.mu, which contains this code, will be created in current directory.

Because it could be regarded as a piece of software I have to include a license to conform with the present legal climate. I choose the *GNU General Public License* [5], please read it before use the code.

---

```
/* We start from the initialisation of variables and constants */

first_slit_is_open := TRUE: // We could separately open the first
second_slit_is_open := TRUE: // or the second slits

grid := 40:     // The number of positions for the particle gun per degree
coverage := 40: // The average number of gun shots per position
num_orbits := 3: // Number of atom's orbits

/* Array of spin variables for each orbit */
spin := array(0..num_orbits, (i)=1 $ i=0..num_orbits):

/* The array stores results of our experiment (initialised with zeros)*/
result := array(−90..90, 1..2, (i,1)=0 $ i=−90..90, (i,2)=0 $ i=−90..90):

/* We need few random generators for:                        */
random_position := random(2*180*grid)/180/grid: // position of a particle gun
```

random_dr := random(2): *// one out of two possible directions of scattering*

/* *We will need a fast solution of equation $x = y + \sin y$ (3) many times*/
interference_grid :=500:
small_step := 2*1.1:
inversion:=**array**(0..interference_grid, (i)=−1 $ i=0..interference_grid):
**for** i **from** 0 **to** 2*interference_grid*small_step+1 **step** 1 **do**
  x := i/interference_grid/small_step*$\pi$:
  inversion[floor((x+sin(x))*interference_grid/2/$\pi$)] := float(x/2/$\pi$):
**end_for**:

/* *Now we start our numerical experiment* */

**for** i **from** 0 **to** 180*grid*coverage **do**
  position := random_position(): *// We choose a position of the gun randomly*
  direction:= random_dr(): *// We choose randomly a direction of scattering*
  /* *There is nothing random beyond this point—everything is deterministic!* */
  slit := floor(position+1):   *// then find which slit is used*
  position := abs(position−1): *// and change the position relative to the centre.*

  **if** ((slit = 1 **and** first_slit_is_open)     *// We let a particle go*
     **or** (slit = 2 **and** second_slit_is_open)) **then** *// through an open slit only*
    x := position*num_orbits: *// The position relative to orbits scales*
    orbit := round(x): *// Which orbit is closest?*

    /* *This is the interference calculations; it is used only if* */
    **if** (spin[orbit] = slit) **then**   *// the spin of the orbit equal to slit*
      position:=frac((floor(x)+inversion[round(frac(x)*interference_grid)])/num_orbits):
      spin[orbit] := 3− slit:  *// And interference flips the spin of the orbit!*
    **end_if**:

    /* *Scattering calculation by the formula $\phi = \pm \arcsin p$ (4).* */
    angle := round((−1)^direction*arcsin(position)/$\pi$*180): *// calculate the angle*
    result[angle,slit] := result[angle,slit]+1: *// and add it to the result*

  **end_if**:
**end_for**:

/* *And finally plot the result* */

res1:=plot::Polygon(plot::Point(i,0,result[i,1]/grid) $ i=−90..90, Color=[0,1,0]):
res2:=plot::Polygon(plot::Point(i,10,result[i,2]/grid) $ i=−90..90, Color=[1,0,0]):
res3:=plot::Polygon(plot::Point(i,20,(result[i,1]+result[i,2])/grid) $ i=−90..90,
                Color=[0,0,1]):
plot(res1,res2,res3);

```
/* The end $*/
```

## Acknowledgments

I am grateful to SciFace ® for granting me and other scientists a free license to use MuPAD [3]. I acknowledge also the usage of free software Lgrind [1] and MetaPost [6]. And should we say once more our thanks to Donald E. Knuth for the wonderful TeX?

School of Mathematics, University of Leeds, Leeds LS2 9JT, UK
*Email*: `kisilv@maths.leeds.ac.uk`
*URL*: `http://maths.leeds.ac.uk/~kisilv/`